\documentclass[conference, letterpaper]{IEEEtran}
\ifCLASSINFOpdf
\else
\fi
\ifCLASSINFOpdf
   \usepackage[pdftex]{graphicx}
\else
\fi

\usepackage[cmex10]{amsmath}
\usepackage{color}
\usepackage{fancyhdr}
\usepackage[caption=false,font=footnotesize]{subfig}

\renewcommand{\thispagestyle}[2]{}

\fancypagestyle{plain}{
        \fancyhead{}
        \fancyhead[C]{first page center header}
        \fancyfoot{}
        \fancyfoot[C]{first page center footer}
}
\begin{document}

%
\title{TIME SERIES ANALYSIS APPLIED TO NOTIFICATIONS OF WORK ACCIDENTS}

\author{\IEEEauthorblockN{Fernandez-Cayo Tony Gabriel}
\IEEEauthorblockA{Faculty of Statistic and Computer Engineering,\\
Universidad Nacional del Altiplano de Puno, P.O. Box 291\\
Puno - Peru.\\
Email: tfernandezc@est.unap.edu.pe}
\and
\IEEEauthorblockN{Torres-Cruz Fred}
\IEEEauthorblockA{Faculty of Statistic and Computer Engineering,\\
Universidad Nacional del Altiplano de Puno,P.O. Box 291\\
Puno - Peru.\\
Email: ftorres@.unap.edu.pe}
}


%


\maketitle

\begin{abstract}
Time series analysis applied to occupational accident reports is a powerful tool for understanding the evolution of occupational accidents over time. It provides valuable information to make informed decisions. In this study, data from reports of work accidents collected from the MINISTRY OF LABOR AND EMPLOYMENT PROMOTION – MTPE were analyzed by time series. Significant patterns and trends in accident reporting have been identified, leading to more effective prevention strategies and better health and safety management.
\end{abstract}

\begin{IEEEkeywords}Time series analysis;Notifications of work accidents;Prevention of occupational hazards;ARIMA models;temporary evolution;Occupational Health and Safety

\end{IEEEkeywords}

\IEEEpeerreviewmaketitle

\section{Introduction}
Nowadays, adequate notification of work accidents is an essential component for the management of occupational safety.\cite{Fernández-MuñizSafetyScience}.
provides valuable information on the occurrence and characteristics of accidents, allowing the identification of patterns, trends and risk factors that may affect safety at work.\par

Time series analysis breaks down data into various components, such as trend, seasonality, and random component, making it easy to identify long-term patterns and seasonal changes in workplace accident notifications. In addition, statistical models such as ARIMA (Autoregressive Integrated Moving Average) models can be used to model and predict the future occurrence of accidents at work... \cite{Rana}.

By understanding the temporal evolution of work accident reports, organizations and those responsible for occupational safety can take more effective preventive measures. Time series analysis provides valuable information to identify periods of increased risk, assess the effectiveness of the preventive measures taken and develop individual strategies to improve safety at work.\cite{Rodríguez}

This article presents a time series approach applied to industrial failure reporting. Data on reported occupational accidents are collected and statistical methods are used to identify important trends and trends. The analysis of time series allows us to better understand the temporal evolution of occupational accidents, contributing to the management of occupational safety and risk prevention in the workplace.\cite{Rodríguez2016}.


\section{METHODS}

In this section, we describe the models used for forecasting.

\subsection{Data collection}
The first step we will do is collect the data, the source from which we obtained the data is the MINISTRY OF LABOR
AND PROMOTION OF EMPLOYMENT - MTPE, SYSTEM OF WORK ACCIDENTS - SAT for notifications of work accidents.
These data include the date and quantity of each notification which are the initials NNAT, NNAM, NNIP, NNEO.\cite{trabajo}.

\begin{figure} [h]
\centering
\includegraphics[width=0.4\textwidth]{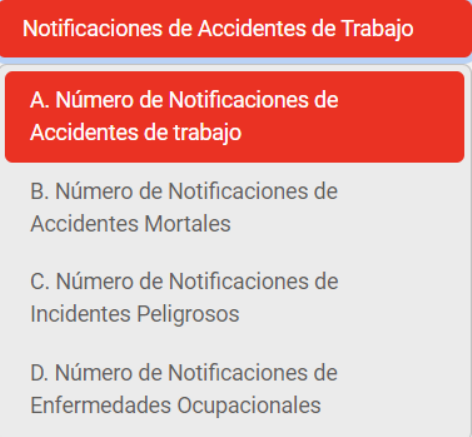}
\caption{\label{fig:fro}data collected from MTC:}
\end{figure}

\subsection{Data preparation}

Once the data is collected, it is important to perform proper cleaning and pre-processing. This involves checking the records for consistency, removing outliers or missing data, and making sure the data is in the proper format for time series analysis. Also, it may be necessary to adjust the temporal frequency of the data.\cite{Cowpertwait}.

\subsection{exploratory visualization}

Before applying analysis techniques, we will perform an exploratory visualization of the data. This involves plotting the time series data on a graph to identify patterns, trends, and seasonality. These visualizations can include line charts, histograms, scatter plots, and time series decomposition. \cite{ Chatfield}

\begin{figure} [h]
\centering
\includegraphics[width=0.4\textwidth]{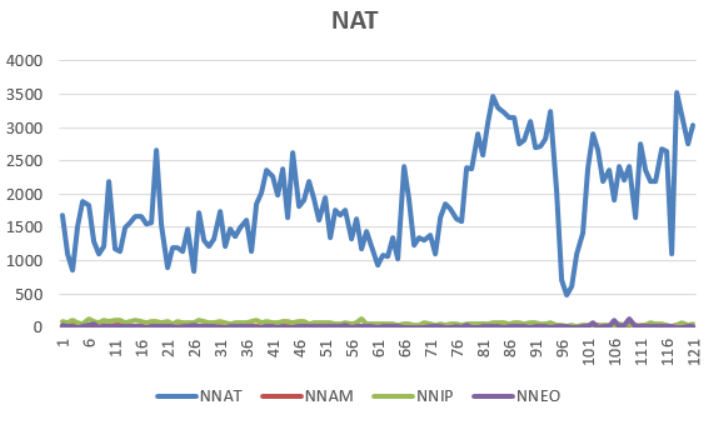}
\caption{\label{fig:fro}you can see the trend over time:}
\end{figure}

\begin{figure} [h]
\centering
\includegraphics[width=0.4\textwidth]{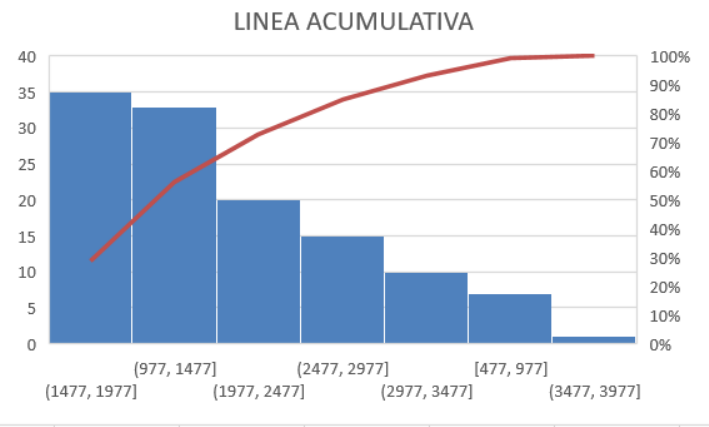}
\caption{\label{fig:fro}our data is sorted decently}
\end{figure}

\subsection{Decomposition of the time series}

Decomposition is an important step in time series analysis. It consists of separating the time series into its main components: trend, seasonality and residual error. The trend shows the general direction of the data in the long term, the seasonality shows repeating patterns in the short term, and the residual error represents random or unexplained variations by the trend and seasonality.\cite{Hyndman}

\begin{figure} [h]
\centering
\includegraphics[width=0.4\textwidth]{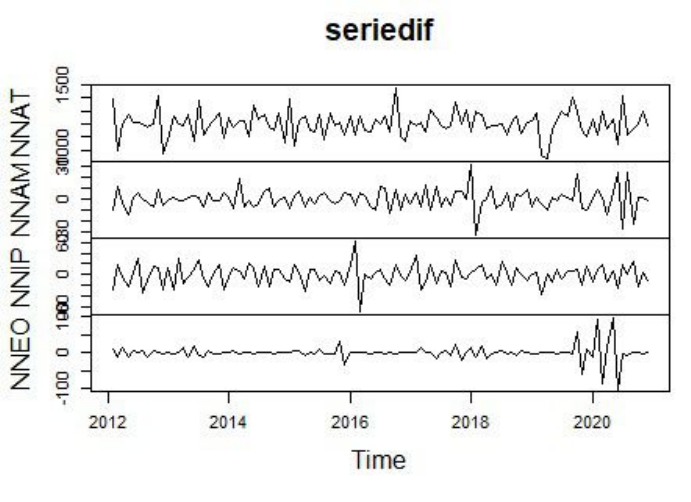}
\caption{\label{fig:fro}sample of the differences of the four data}
\end{figure}

\begin{figure} [h]
\centering
\includegraphics[width=0.4\textwidth]{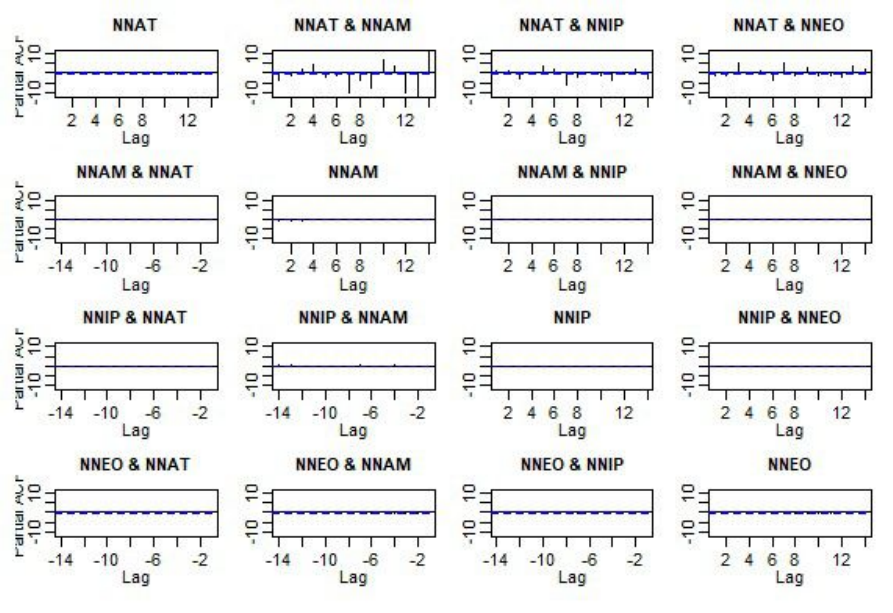}
\end{figure}

\begin{figure} [h]
\centering
\includegraphics[width=0.4\textwidth]{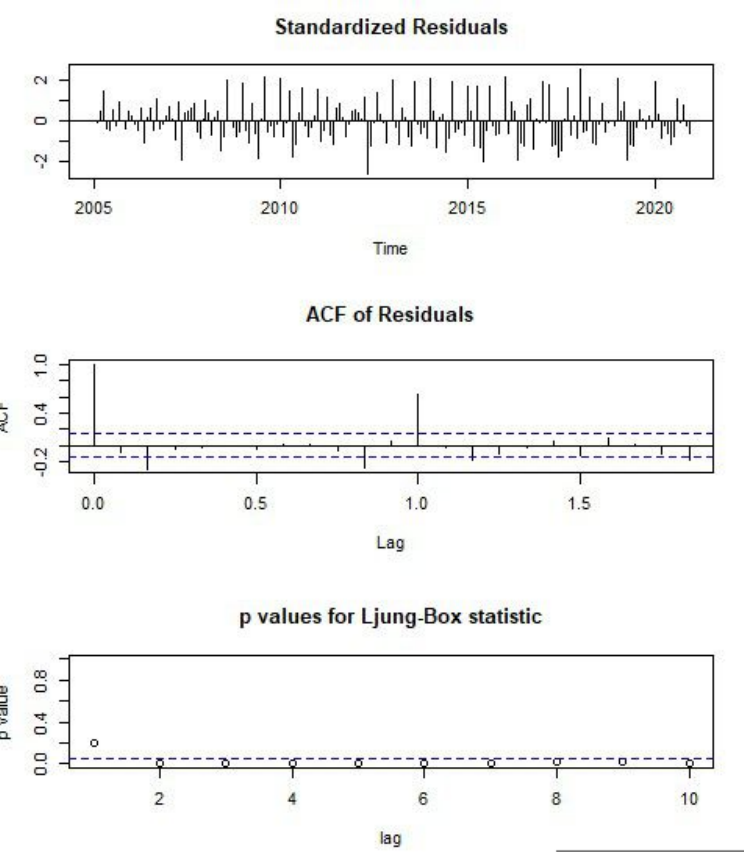}
\caption{\label{fig:fro}we make an autocorrelation}
\end{figure}

\subsection{Modeling and forecast}

Once the time series is decomposed, we will use different statistical models to forecast future work accidents. Some common techniques include ARIMA (Autoregressive Integrated Moving Average) models, exponential smoothing models, and regression models. These models allow us to predict the occurrence of accidents and evaluate the impact of possible predictor variables on the time series.\cite{Brockwell} 

\begin{figure} [h]
\centering
\includegraphics[width=0.5\textwidth]{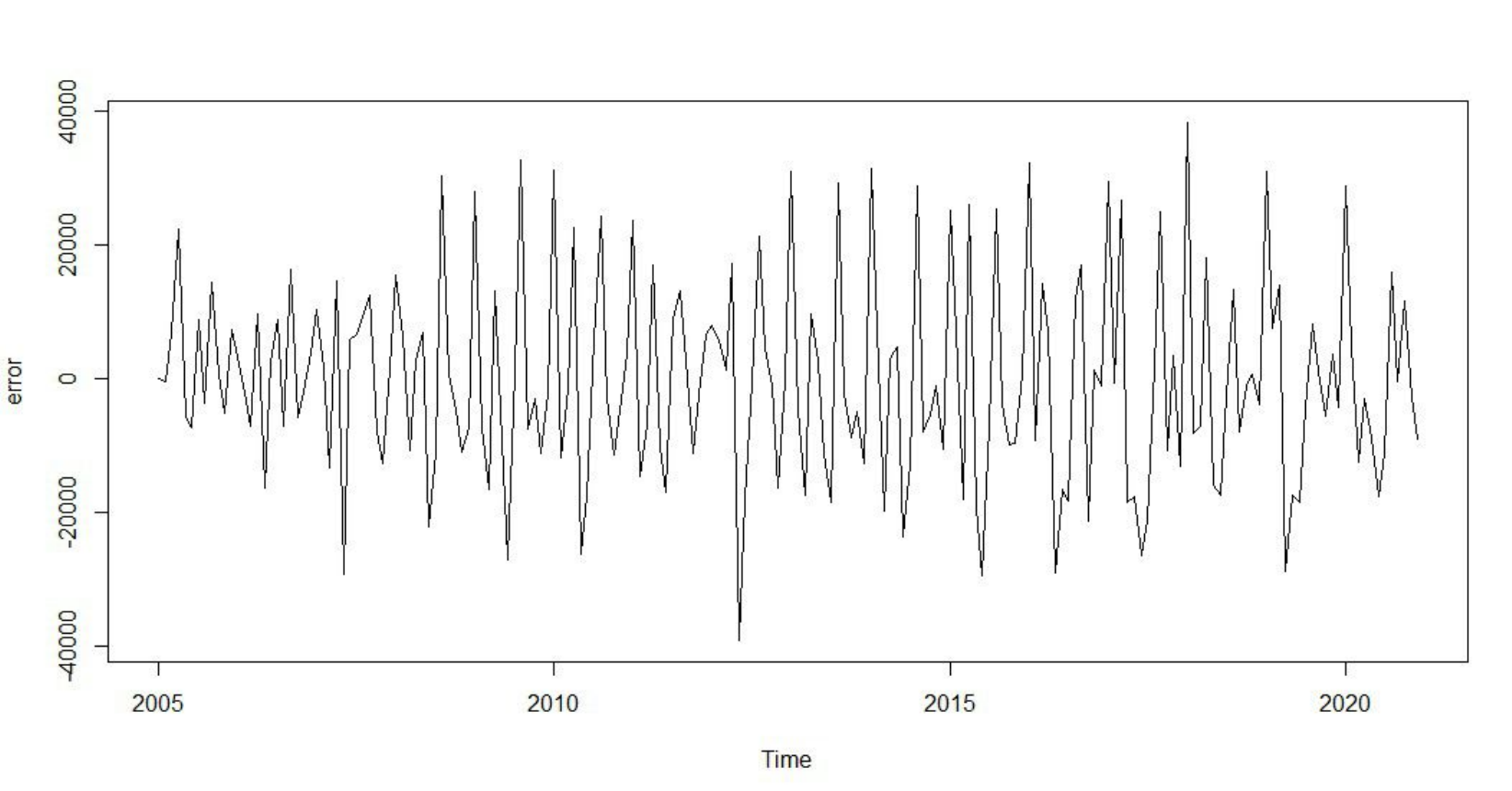}
\caption{\label{fig:fro}We can visualize the series by graphing it}
\end{figure}

\subsection{Model evaluation}

Evaluates the accuracy of the forecast model using error measures such as the mean square error (MSE) or the mean absolute error (MAE). Compare the predicted values with the actual values of workplace accident notifications to determine the effectiveness of the model.It is a stationary Gaussian process where the mean and variance are independent and the covariance of two variables will depend on the time lag of k
\cite{Cowpertwait}

\section{RESULTS}

Studies conducted from 2012 to 2020 We identified time series patterns and trends revealed significant patterns and trends in the occurrence of work accidents over time. Seasonal fluctuations were observed, with increases in accidents during certain periods of the year. In addition, cyclical patterns were identified in which accidents tend to recur at certain times of the day or days of the month.\cite{Cowpertwait} These results provide key information to direct preventive measures at times and areas of greatest riska. Time series analysis revealed significant seasonal patterns in accident reports with increases in certain months of the year. There was also an overall downward trend in accidents over the five years of the study. In addition, drastic changes were identified in the accident report in response to the security measures of the institution.\cite{Shumway}

\begin{figure} [h]
\centering
\includegraphics[width=0.5\textwidth]{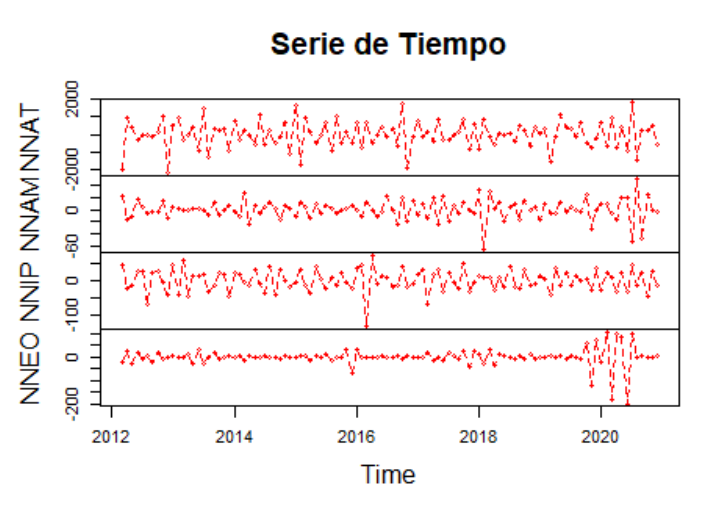}
\caption{\label{fig:fro}Here it can be seen that NNAT and NNEO have a big difference, they have more deaths:}
\end{figure}

Through time series analysis, forecast models were developed to predict the future occurrence of work accidents.\cite{Wang} These models make it possible to anticipate periods of increased risk and take proactive preventive measures. For example, if an upward trend in the occurrence of accidents has been identified, organizations can intensify security measures during those specific periods to reduce the probability of incidents.\cite{stoffer}

\begin{figure} [h]
\centering
\includegraphics[width=0.5\textwidth]{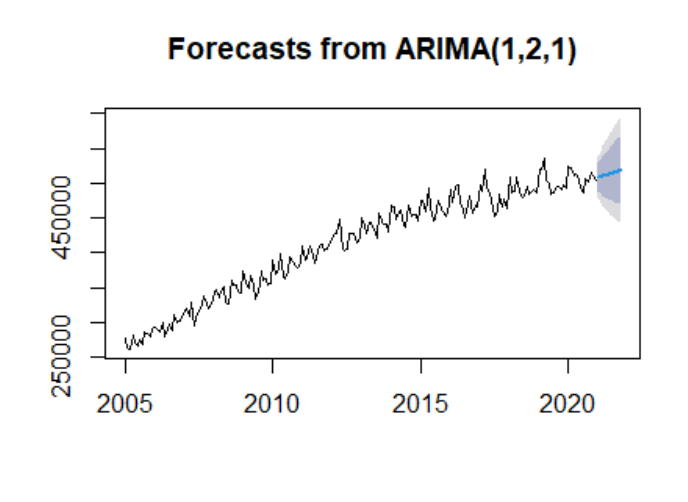}
\caption{\label{fig:fro}As can be seen in the graph, our data is in ascending order and has variance. This means that accident notifications increase every year:}
\end{figure}
\vspace{3mm}

\begin{figure} [h]
\centering
\includegraphics[width=0.5\textwidth]{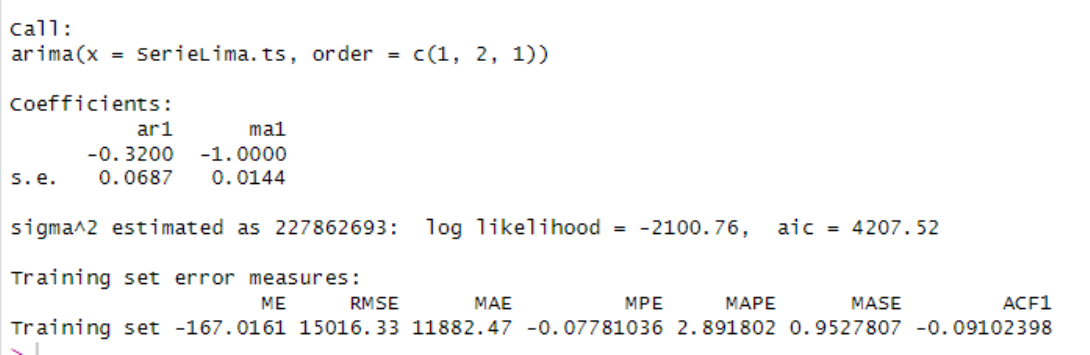}
\caption{\label{fig:fro}here we make a summary to see the data:}
\end{figure}
\vspace{3mm}

\begin{figure} [h]
\centering
\includegraphics[width=0.5\textwidth]{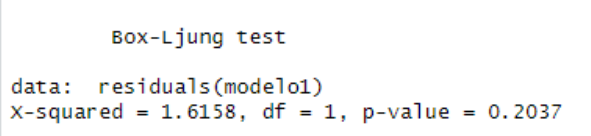}
\caption{\label{fig:fro}here we make a box and it can be seen that it gives us p-value 0.2037:}
\end{figure}
\vspace{3mm}
\begin{figure} [h]
\centering
\includegraphics[width=0.5\textwidth]{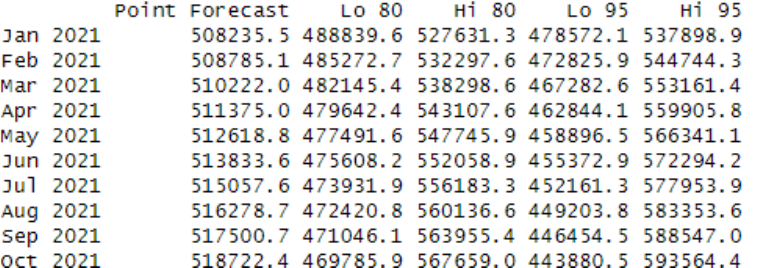}
\caption{\label{fig:fro}Here you can see the forecast for the year 2021 with forecast:}
\end{figure}

\vspace{2mm}
\section{Discussion}

The results of the time series analysis in the notifications of work accidents provide a solid base to improve the existing preventive measures.\cite{Wei} By understanding patterns and trends, organizations can identify recurring risk areas and focus resources and efforts on implementing more effective preventative measures. These findings support informed decision making to improve job security.\cite{box}The results indicate the importance of considering seasonal factors in accident reporting and adapting preventive measures accordingly. The general downward trend in accidents suggests that the safety actions implemented have been effective. However, continuous monitoring and evaluation of prevention strategies is required to address the identified changes in accident reporting.\cite{Chowdhury}

\vspace{5mm}

The results of the time series analysis are also useful for resource planning and job security strategies. By predicting the future occurrence of accidents, organizations can allocate resources more efficiently and plan preventative measures at times and areas of greatest risk. This contributes to a more effective use of resources and a reduction in the costs associated with work accidents.\cite{Hämäläinen}

\vspace{3mm}

\section{Conclusions}
\vspace{3mm}

Time series analysis of work accident notifications provides valuable results to improve occupational safety and prevent risks in the work environment. The patterns, trends, and changes identified through this analysis make it possible to focus preventive measures, evaluate the effectiveness of implemented interventions, and plan data-based strategies to protect workers. By using time series analysis, organizations can make more informed and effective decisions in their pursuit of a safer and healthier work environment.\cite{visuri}Time series analysis of workplace accident reporting provides deeper insight into patterns and trends in workplace safety. These findings can be used to improve risk prevention strategies and safety management in the work environment. Ongoing collaboration between employers, workers and competent authorities is needed to ensure a safe and healthy working environment.\cite{Cowpertwaitt}

%

\end{document}